\documentclass[square]{ws-procs975x65}

\begin{document}
\def\beq{\begin{equation}}
\def\eeq{\end{equation}}

\newcommand{\bea}{\begin{eqnarray}}
\newcommand{\eea}{\end{eqnarray}}

\def\Tdot#1{{{#1}^{\hbox{.}}}}
\def\Tddot#1{{{#1}^{\hbox{..}}}}

\newcommand{\dn}[2]{{\mathrm{d}^{{#1}}{{#2}}}}
\def\PX{P_{,X}}
\def\s{\sigma}

\def\tX{{\tilde X}}
\def\tG{{\tilde G}}
\def\tP{{\tilde P}}

\def\h{{\cal H}}

\def\d{\delta}
\def\B{{\bar B}}
\def\E{{\bar E}}
\def\V{{\cal V}}
\def\P{{\cal P}}
\def\R{{\cal R}}
\def\D{\nabla}

\def\L{{\rm L}}
\def\k{{\vec k}}
\def\x{{\vec x}}

\def\mP{M_P}

\def\bfphi{{\bf \phi}}

\def\bx{{\bf x}}
\def\bk{{\bf k}}
\def\PP{{\cal P}}

\title{Glimpses into the early Universe}

\author{David Langlois}

\address{APC (Astroparticules et Cosmologie), CNRS-Universit\'e Paris 7\\
10, rue Alice Domon et L\'eonie Duquet, 75205 Paris Cedex 13, France\\
E-mail: langlois@apc.univ-paris7.fr}

\begin{abstract}
This contribution gives a brief overview of the theoretical ideas underlying our
current understanding of the early Universe. Confronting the predictions of the early Universe models 
with cosmological observations, in particular of the cosmic microwave
background fluctuations,  will improve our knowledge about the physics of the primordial Universe.

\end{abstract}


\bodymatter

\section{Introduction}
Inflation is today the main theoretical framework to describe  the early Universe. In thirty years of existence,  inflation, in contrast with earlier competitors,   has survived the confrontation with  cosmological data, which have tremendously improved over the years.  Indeed, the fluctuations of the Cosmic Microwave Background (CMB) had not yet been measured when inflation was invented, whereas they give us today  a remarkable   picture of  the cosmological perturbations in the early Universe.  In the future, one can hope that even more precise observations will allow us to test  inflation further, and also to discriminate between the many different possible realizations of inflation. 

This contribution discusses the basic ideas underlying inflation (many more details  can be found in e.g. \cite{cargese}) and some more recent results.

\section{Cosmological evolution}
Modern cosmology is based on the theory of general relativity, according to which our Universe is described by a four-dimensional 
geometry $g_{\mu\nu}$ that  satisfies Einstein's equations
\beq
\label{einstein}
G_{\mu\nu}\equiv R_{\mu\nu}-\frac12 R \, g_{\mu\nu}=8\pi G \, T_{\mu\nu},
\eeq
where $R_{\mu\nu}$ is the Ricci tensor, $R\equiv g^{\mu\nu}R_{\mu\nu}$ the scalar curvature and $T_{\mu\nu}$ the energy-momentum tensor of the matter distribution. 

The basic assumption of cosmology, which has  
been confirmed by observations so far,  
is to consider, as a first approximation, the universe as 
being homogeneous and 
isotropic. This leads to  the FLRW (Friedmann-Lema\^itre-Robertson-Walker) spacetimes, with  metric
\beq
ds^2=-dt^2+a^2(t)\left[{dr^2\over{1-\kappa r^2}}+
r^2\left(d\theta^2+\sin^2\theta\,  d\phi^2\right)\right],
\label{RW}
\eeq
where $\kappa=0,-1,1$ determines the curvature of spatial 
hypersurfaces: respectively flat, elliptic or hyperbolic.
Moreover, the  matter content compatible with  homogeneity
and isotropy is necessarily characterized by an energy-momentum tensor of the 
form 
\beq
\label{T}
T^{\mu}_{\  \nu}={\rm Diag}\left[-\rho(t), P(t),  P(t), P(t)\right]\, 
\eeq
where $\rho$ corresponds to an energy density and $P$ to a pressure.

Substituting the  metric (\ref{RW})  and the energy-momentum tensor (\ref{T}) 
into Einstein's equations (\ref{einstein}) gives the  
Friedmann equations\index{Friedmann equations},
\begin{eqnarray}
H^2\equiv\left({\dot a\over a}\right)^2 = {8\pi G \over 3}\rho- {\kappa\over a^2},\qquad
{\ddot a\over a} = -{4\pi G\over 3}\left(\rho+3 P\right)\, ,
\label{friedmann}
\end{eqnarray}
which govern the time evolution of the scale factor $a(t)$. 
There are several types of matter in the Universe, in particular pressureless matter (baryonic matter and the mysterious dark matter) and a gas of cosmological photons,  characterized by a temperature $T$, which scales like $1/a(t)$. Going backwards in time, radiation dominates, with  higher and higher temperatures in the early Universe. 

From an observational point of view, the two most important events in cosmological history are : i) nucleosynthesis ($T\sim 0.1$ MeV),  when the lightest nuclei were formed;  ii) last scattering ($T\sim 3000$ K), when the Universe became quasi-transparent (due to the sudden suppression of interactions between photons and matter, as nuclei and electrons combined into neutral atoms). Photons that were emitted at that epoch are observed  in the CMB radiation, discovered in 1964 by Penzias and Wilson. In 1992, the COBE satellite detected its anisotropies at a level of $10^{ -5}$. 
Since then, these fluctuations have been measured with increasing precision, lately by the WMAP satellite and in the near future, by the Planck satellite. One of the main goals of primordial cosmology is 
to explain the origin of these primordial fluctuations.

\section{Inflation}
Inflation is a phase of accelerated expansion, i.e. $\ddot a>0$, in the early Universe. 
Initially, inflation was introduced to provide an explanation for several puzzles of the standard hot Big Bang model, in particular the flatness problem (or why the present spatial geometry is so close to  Euclidean geometry) and the horizon problem (or why the CMB sky is so homogeneous on scales larger than the causal horizon at the time of last scattering, as defined in the standard Big Bang model). But it was realized, soon after, that inflation also gives a very natural explanation for the origin of primordial perturbations: they simply arise from  quantum vacuum fluctuations that got amplified when their wavelength, proportional to the scale factor, is stretched out beyond the Hubble radius $H^{-1}$.

The simplest way to get inflation is to assume the existence of a scalar field, governed by the 
action 
\beq
\label{action_scalar_field}
S_\phi=\int d^4x\sqrt{-g}\left(-{1\over 2}\partial^\mu\phi\partial_\mu\phi
-V(\phi)\right)\, ,
\eeq
where $g\equiv {\rm det} (g_{\mu\nu})$ and $V(\phi)$ is the potential of the scalar field. In a FLRW spacetime, the energy density and pressure are respectively given by 
\beq
\rho={1\over 2}\dot\phi^2+V(\phi), \qquad
P={1\over 2}\dot\phi^2-V(\phi)\, .
\eeq
Whenever the kinetic energy is negligible with respect to the potential energy, the equation of state is effectively $P\simeq -\rho$, which leads to an acceleration, according to the second Friedmann equation in (\ref{friedmann}). 
This can happen in the so-called slow-roll regime, for sufficiently flat potentials.

\section{Cosmological perturbations}
Let us now consider  the perturbations during inflation. In addition to the scalar field perturbation $\delta\phi$, 
one must take into account the metric perturbations as well and introduce the (scalarly) perturbed metric
\beq 
ds^2= -(1+2A)dt^2+ 2 a(t) \partial_iB\, dx^idt+
a^2(t)\left[(1-2\psi)\delta_{ij}
+2\partial_i\partial_jE\right]dx^idx^j\, .
\eeq
Using Einstein's equations and  coordinate freedom, it can be shown that there is in fact a single dynamical scalar degree of freedom,
\beq
Q\equiv \delta\phi+\frac{\dot\phi}{H}\psi\equiv \frac{\dot\phi}{H}\R \, .
\eeq
It can be seen either as a pure scalar field perturbation (in a coordinate system such that $\psi=0$) or as a pure metric perturbation $\R$ (up to the factor $\dot\phi/H$) if one chooses  the uniform scalar field hypersurfaces as constant time hypersurfaces. 
Its dynamics is governed by the action
\beq
S={1\over 2}\int d\tau \, \, d^3x\, \left[{u'}^2+\partial_i u \partial ^i u
+{z''\over z}u^2\right],\qquad z\equiv a\frac{\dot\phi}{H}
\eeq
where it is convenient to use the new variable $u\equiv a Q$ and the conformal time $\tau$ ($d\tau=dt/a$).

Let us now quantize $u$ by following the standard procedure of quantum field theory.
One  treats $u$ as a quantum field denoted $\hat u$, which can be
expanded in Fourier space as 
\index{quantization}
\beq
\label{Fourier_quantum}
\hat u (\tau, \vec x)={1\over (2\pi)^{3/2}}\int d^3k \left\{{\hat a}_{\vec k} u_k(\tau) e^{i \vec k.\vec x}
+ {\hat a}_{\vec k}^\dagger u_k^*(\tau) e^{-i \vec k.\vec x} \right\},
\eeq
where  the $\hat a_{\vec k}^\dagger$ and  $\hat a_{\vec k}$ are 
 creation and annihilation operators that  satisfy the 
usual commutation rules.
In the slow-roll regime, the expansion is  quasi-de Sitter  (i.e. such that $\dot H\ll H^2$) with $a\simeq -1/H\tau$, and $z''/z\simeq a''/a\simeq 2/\tau^2$ (where $\tau$ grows from $-\infty$ to $0$). 
The most natural choice for the mode function $u_k(\tau)$  is the particular solution (of the  classical equation of motion) 
\beq
u_k=\sqrt{\hbar\over 2k}e^{-ik\tau }\left(1-{i\over k\tau}\right), 
\label{u_k}
\eeq 
which means that each Fourier mode is initially in the usual Minkowski vacuum, when $k|\tau|\gg 1$, i.e. when its wavelength is smaller than the Hubble radius. Later, its wavelength is stretched on super Hubble scales, i.e. $k|\tau|\ll 1$, and the mode undergoes an amplification (the last term in the parentheses blows up). The resulting power spectrum (i.e. the Fourier transform of the correlation function) is
\beq
\P_Q\equiv \frac{k^3}{2\pi^2}\frac{|u_k|^2}{a^2}\simeq \hbar \left({H\over 2\pi}\right)^2 \, ,
\qquad (k\ll aH)
\eeq
which corresponds to a curvature fluctuation $\P_\R^{1/2}= H^2/(2\pi \dot\phi)$ (with $\hbar=1$). 

It is easy to relate these fluctuations generated {\it during} inflation to the fluctuations of ordinary matter in the subsequent radiation and matter dominated eras, by invoking a conservation law. 
Indeed, the conservation of the energy-momentum tensor 
for any perfect fluid, characterized by the energy density $\rho$, the pressure $p$ and the four-velocity $u^a$, leads 
to the {\it exact} relation\cite{Langlois:2005ii,Langlois:2005qp}
\beq
\label{dot_zeta}
\dot\zeta_a\equiv {\cal L}_u\zeta_a=
-{\Theta\over{3(\rho+p)}}\left( \nabla_a p -
\frac{\dot p}{\dot \rho} \nabla_a\rho\right), 
\eeq
where we have defined 
\beq
 \zeta_a\equiv 
\nabla_a\alpha-\frac{\dot\alpha}{\dot\rho}\nabla_a\rho, \quad \Theta=\nabla_a u^a, \quad \alpha=\frac{1}{3}\int d\tau \,
\Theta,
\eeq
and where a dot denotes a Lie derivative with respect to $u^a$ (which reduces to a derivative along $u^a$ for scalars,  e.g. $\dot\rho\equiv u^a\nabla_a\rho$).  The quantity $e^\alpha$ can be interpreted as an inhomogeneous generalization of the scale factor, as defined by an observer following the fluid. For linear perturbations, the identity (\ref{dot_zeta}) implies that the quantity
\beq
\zeta\equiv-\psi-\frac{H}{\dot\rho}\delta\rho
\eeq
is conserved on super-Hubble scales ($k\ll aH$) for adiabatic perturbations, i.e. such that $\delta p- (\dot p/\dot\rho)\delta\rho=0$. Moreover, it can be shown that $-\zeta$ coincides with $\R$ on super-Hubble scales. For single field inflation, this conservation law holds and the curvature perturbation remains unchanged until the perturbation reenters  the Hubble radius, much later during the radiation era.

\section{Beyond the simplest models}
So far, the simplest models of inflation are compatible with observational data (see \cite{WMAP}) but it is worth studying more refined models for at least two reasons. First, models inspired by high energy physics are usually more complicated than the simplest phenomenological inflationary models. Second, exploring  larger classes of inflation models and identifying their specific observational features is a useful  preparation for the interpretation of  future data. At present, two types of extensions beyond the simplest scenarios have been mainly studied: models with non standard kinetic terms and/or  with multiple  scalar fields.
  
Scenarios involving several scalar fields include  models with {\it multiple inflatons},  where several scalar fields affect directly the  inflationary evolution, but also models where the extra scalar field(s) plays a r\^ole only later. 
In all cases, the crucial novelty is the generation during inflation of extra perturbations, usually called entropy modes, in addition to the adiabatic mode (corresponding to fluctuations along the inflation trajectory). The entropy fluctuations can be transferred into the final curvature perturbation, during or after inflation. This means that the quantities $\R$ or $\zeta$ are {\it a priori} no longer conserved in a multi-field set-up, as first pointed out in \cite{sy} (see also \cite{Lalak:2007vi}). 

In some of these more sophisticated models, the primordial perturbations exhibit non-Gaussianities that could be detectable in future observations, whereas the simplest single field models predict an undetectable level of non-Gaussianities. In this context, an important observable is the three-point function or its Fourier transform, often written as 
\beq
  \langle  \zeta_{\bk_1} \zeta_{\bk_2} \zeta_{\bk_3} 
\rangle
\, \equiv 
\frac{6}{5}
f_{\rm NL} (k_1,k_2,k_3)\left[P_\zeta(k_1)P_\zeta(k_2)+ 2 \, {\rm perms}\,  \right](2 \pi)^3
\delta^{(3)}(\sum_i \bk_i) \,
\eeq
where $P_\zeta(k)$ is defined by $ \langle  \zeta_{\bk_1} \zeta_{\bk_2}\rangle=P_\zeta(k)(2 \pi)^3 \delta^{(3)}(\bk_1+\bk_2)$. The three-point function can be computed
 for any model of inflation by using the (non-linear) relation between the curvature perturbation $\zeta$ and the various scalar field fluctuations generated during inflation.  Intrinsic non-Gaussianities of the scalar fields, i.e. non-vanishing three-point functions for the scalar field perturbations, lead to non-Gaussianities of {\it equilateral} shape (where the signal peaks at $k_1\sim k_2\sim k_3$) whereas a non-linear (classical) relation between $\zeta$ and the scalar field fluctuations lead to non-Gaussianities of {\it local} shape (which peaks at $k_1\ll k_2, k_3$).

An analysis of   general  multi-field models with  an action of the form
\beq
\label{P}
S =  \int d^4 x \sqrt{-g}\left[\frac{R}{16\pi G}   +   P(X^{IJ},\phi^K)\right], \qquad  X^{IJ}=-\frac12 \nabla_\mu \phi^I  \nabla^\mu \phi^J\, ,
\eeq
where $P$ is an arbitrary function, 
 can be found in \cite{lrst08b} (see also \cite{lr08} for a more restricted class of models).

An illustrative example combining both  non-Gaussianities and multi-field effects is Dirac-Born-Infeld (DBI) inflation~\cite{st,ast}, which arises from the motion of a  
$D3$-brane in an internal  six-dimensional compact space, as one can encounter in the context of  string theory compactifications.  In this case, the function $P$
is given by
\beq
P=-\frac{1}{f(\phi^I)}
\left[\sqrt{{\rm det}\left(\delta^\mu_\nu+f G_{IJ}\partial^\mu\phi^I\partial_\nu^J\right)}-1\right]-V(\phi^I),
\eeq
where the scalar fields $\phi^I$ are directly related to the compact coordinates of the brane, and where the functions $f$ and $G_{IJ}$, which depends on the $\phi^I$, are determined by the 
10-dimensional geometry (bulk forms can also be included~\cite{lrs09}).

During multi-field DBI inflation, both adiabatic and entropic modes can be generated, with  power spectra $\P_{Q_\sigma}= (H/2\pi)^2$ and 
$\P_{Q_s}= (H/2\pi c_s)^2$ respectively~\cite{lrst08a}.
The entropy power spectrum is  thus enhanced by its dependence on the sound speed $c_s=\sqrt{1-f\dot\sigma^2} \,  <1$, and  the final curvature is
\beq
\R=\frac{H}{\dot \s}\left(Q_{\sigma}+T_{ {\cal R}  {\cal S} } c_sQ_{s}\right), \qquad \dot\sigma^2\equiv G_{IJ}\dot\phi^I\dot\phi^J, 
\eeq
where $T_{ {\cal R}  {\cal S} }$ parametrizes the transfer from entropy modes into the curvature perturbation.

By expanding the action beyond the second order, one can also determine the non-Gaussianities 
generated in this class of models~\cite{lrst08a}. This leads to a bispectrum of equilateral shape with
\beq
f_{NL}^{(3)}=-\frac{35}{108}\frac{1}{c_s^2}\frac{1}{1+T^2_{{\cal R} {\cal S}} },
\eeq 
which corresponds to the single-field result but with a suppression due to the entropy-curvature transfer. At the next order, in the trispectrum, 
multi-field effects induce a shape of non-Gaussianities that differs from the single-field case~\cite{Mizuno:2009cv}.

\end{document}